\documentclass[12pt]{article}

\usepackage{graphicx}

\newcommand{\be}{\begin{equation}}
\newcommand{\ee}{\end{equation}}
\newcommand{\bea}{\begin{eqnarray}}
\newcommand{\eea}{\end{eqnarray}}

\newcommand{\tA}{\tilde A}
\newcommand{\tB}{\tilde B}

\newcommand{\RR}{\rangle}
\newcommand{\LL}{\langle}

 \makeatletter
\newdimen\normalarrayskip
\newdimen\minarrayskip
\normalarrayskip\baselineskip
\minarrayskip\jot
\newif\ifold             \oldtrue

\makeatother
\newlength{\extraspace}
\newlength{\extraspaces}
\begin{document}

\addtolength{\baselineskip}{.4mm}

\thispagestyle{empty}

\begin{flushright}
\baselineskip=12pt
$         $\\
\hfill{  }\\
\end{flushright}
\vspace{.5cm}

\begin{center}
\baselineskip=24pt

{\LARGE {Nonlinear Qubit Transformations}}\\[15mm]

\baselineskip=12pt

{Lucien Hardy and David D. Song} \\[%
8mm]
{Centre for Quantum Computation\\[0pt]
Clarendon Laboratory, Department of Physics \\
University of Oxford, Parks Road, Oxford OX1 3PU, U.K.} \vspace{2cm}

{\bf Abstract}

\begin{minipage}{15cm}
\baselineskip-12pt  We generalise our previous results of
universal linear manipulations [Phys. Rev. A {\bf 63}, 032304
(2001)] to investigate three types of nonlinear qubit
transformations using measurement and quantum based schemes.
Firstly, nonlinear rotations are studied.  We rotate different
parts of a Bloch sphere in opposite directions about the $z$-axis.
The second transformation is a map which sends a qubit to its
orthogonal state (which we define as ORTHOG).  We consider the
case when the ORTHOG is applied to only a partial area of a Bloch
sphere. We also study nonlinear general transformation, i.e.
$(\vartheta,\varphi)\rightarrow (\vartheta-\alpha,\varphi)$,
again, applied only to part of the Bloch sphere. In order to
achieve these three operations, we consider different measurement
preparations and derive the optimal average (instead of universal)
quantum unitary transformations. We also introduce a simple method
for a qubit measurement and its application to other cases.

\end{minipage}
\end{center}
\vspace{2cm}
\section{Introduction}
In recent years interests in the field of quantum computation has
grown rapidly (see \cite{qc1,qc2,qc3} for reviews). For a deeper
understanding of the fundamental principles and limitations of
quantum computing, it is important to investigate optimal
operations for unknown states. In a Bloch sphere notation, an
unknown qubit has the following form, \be |\psi
(\vartheta,\varphi)\RR =\cos\frac{\vartheta}{2} |0\RR +
e^{i\varphi} \sin\frac{\vartheta}{2} |1\RR \label{bloch}\ee There
have been a number of studies to find optimal manipulations of
quantum states. Bu\v{z}ek and Hillery introduced
\cite{clon1,clon2} a universal quantum cloner such that, given a
single unknown qubit, it produces two approximate copies of the
input qubit. Subsequently, Gisin and Massar generalised it
\cite{gisin0} to a $N$ to $M$ cloner. These cloners have been
proven to be optimal by various researchers
\cite{werner,bruss,fuchs,gisin0}. A universal quantum entangler
has also been introduced in \cite{entangler}. It entangles two
qubits where one is unknown while the other is a known reference
state. Moreover, Mor and Terno introduced a universal quantum
disentangler \cite{disent1,disent2,disent3} (also see
\cite{ghosh}). They considered a bipartite state and its
transformation into a product state of the reduced density
matrices of the two subsystems. Bu\v{z}ek and Hillery considered
\cite{hillery} a slight different disentangler which extracts an
unknown qubit entangled with a reference state. Recently,
universal-NOT (U-NOT) transformation has been studied in
\cite{buzek1,buzek2,gisin}.  The perfect operation of $|\psi\RR
\rightarrow |\psi^{\perp}\RR$ is not allowed due to its
anti-unitarity. In a single qubit case, the fidelity of imitating
such operation was shown to be 2/3. This is same as the
measurement fidelity which is the optimal efficiency of measuring
an unknown qubit. Universality of these transformations imply that
the operations yield the same fidelity regardless of the input
states. In \cite{song1}, more general universal manipulations of a
qubit were derived by the present authors. We considered a linear
transformation, \be |\psi (\vartheta , \varphi)\RR \rightarrow
|\psi (\vartheta - \alpha, \varphi - \beta)\RR \label{PRtran}\ee
where the transformation of $\varphi$ by $\beta$ can always be
done by unitary rotation about the $z$-axis.  In performing the
transformation of $\vartheta$, there are two types of approaches
we could take. One, which we call the {\it {measurement scheme}},
is to measure the qubit then prepare another qubit according to
the transformation in (\ref{PRtran}). The other approach, which we
call the {\it {quantum scheme}}, is by a quantum unitary
transformation on the system plus an ancilla. In case of the
quantum approach, we found the procedures of unitary
transformation that optimise fidelity fall into two classes
depending on the phase angle change $\alpha$. For $0 \leq \alpha
\leq \pi/2$ the best way is simply an identity map while for
$\pi/2 \leq \alpha \leq \pi$, Bu\v{z}ek et al.'s U-NOT gate is the
optimal transformation.

The transformations we will consider in this paper do not respect
the symmetry of the Bloch sphere and therefore it does not make
sense to impose universality.  Hence, we will consider
optimisation of the average fidelity without imposing that the
fidelity be the same for each input state.

We also want to define an operation which sends a qubit to its
orthogonal state, \be {\rm ORTHOG}:|\psi\RR \rightarrow
|\psi^{\perp}\RR \ee  When we impose universality on the operation
ORTHOG, it would correspond to the Bu\v{z}ek et al.'s U-NOT gate.

In this paper, we attempt to generalise the linear transformation
in (\ref{PRtran}) to various nonlinear cases.  Gisin has proposed
\cite{gisin1} a rather peculiar rotation where upper and lower
hemispheres of a Bloch sphere are rotated in opposite directions
about the $z$-axis.  We study both measurement and quantum
approaches of this type of nonlinear map and also consider other
types of nonlinear operations.  Our results are summarised as
follows;
\begin{itemize}
\item We study a nonlinear transformation as shown in figure
\ref{rotationPI2}, which rotates different parts of a Bloch sphere
in opposite directions about the $z$-axis.  When the two parts are
upper and lower hemispheres, the optimal map is either identity or
the unitary rotation by $\pi$. We find the average optimal
transformations for general nonlinear rotations.
\item Nonlinear ORTHOG gates are studied. We investigate averaged
(instead of universal) ORTHOG transformations that are applied
only to a partial area of a Bloch sphere as shown in figure
\ref{Unot}. Bu\v{z}ek et al.'s U-NOT gate appears as a one
particular case when the transformation is applied to the whole
Bloch sphere and universality is imposed. In general, we show the
optimal map is either identity or the unitary operation
$\sigma_x=\left(
\begin{array}{cc} 0&1
\\ 1 & 0\end{array} \right) $, a Pauli matrix.
\item We consider nonlinear general transformations (i.e.
$(\vartheta,\varphi)\rightarrow (\vartheta-\alpha,\varphi)$) as
shown in figure \ref{arbit}.  Unlike the ORTHOG-gate, the averaged
general transformation has higher fidelity than the universal
manipulation of a qubit in general.  We show the optimal
operations for different values of $\alpha$.
\item We also show a simple way of measuring an unknown qubit,
measurement-based U-NOT and general linear transformation while
obtaining the same fidelities as in the case of a conventional
method.
\end{itemize}
In sect. 2, we review a traditional method of qubit measurement
and show a simpler way to achieve the same fidelity.  This
particular way also works equally well for measurement based U-NOT
and the general linear transformation in (\ref{PRtran}).  In sect.
3, we study Gisin's proposed nonlinear rotations. In sect. 4 and
5, we study nonlinear transformation of ORTHOG and the general
transformation in (\ref{PRtran}) where the operations are applied
only to chosen areas of a Bloch sphere.

\section{Measuring an Unknown Qubit}
In order to get maximum information about a single unknown qubit,
we measure the qubit along any chosen basis $\{ |\phi\RR ,
|\phi^{\perp}\RR\}$ \cite{mea1,mea2}. If the result is $|\phi\RR$,
then we guess the qubit to be $|\phi\RR$ and if the result is
$|\phi^{\perp}\RR$, then we guess $|\phi^{\perp}\RR$. In density
matrix terms, the state prepared is written as follows, \be \rho_1
= |\LL \psi|\phi\RR |^2 |\phi\RR\LL \phi| + |\LL
\psi|\phi^{\perp}\RR|^2 |\phi^{\perp}\RR\LL \phi^{\perp}|
\label{Rho1}\ee Averaging over all possible $|\phi\RR$'s (assuming
a uniform distribution over the Bloch sphere), we obtain
$\overline{\rho_1}$.  In order to obtain the fidelity, we take
another integral of $\LL \psi|\overline{\rho_1}|\psi\RR$ over
possible inputs of $|\psi\RR$. After the integration, the fidelity
turns out to be 2/3.

A simpler way of achieving the same result is as follows. Instead
of (\ref{Rho1}), we measure onto the $\{ |0\RR,|1\RR\}$ basis.  As
in the case of (\ref{Rho1}), if we obtain $|0\RR$ then we guess
the unknown state to be $|0\RR$ and when $|1\RR$ is obtained  we
guess $|1\RR$.  In a density matrix form, it is  \be \sigma_1 =
|\LL \psi|0\RR |^2 |0\RR\LL 0| + |\LL\psi |1\RR|^2 |1\RR\LL 1|
\label{Rho2}\ee  We don't need to take the average of $\sigma_1$
since we chose one particular basis $\{ |0\RR,|1\RR\}$ instead of
arbitrary basis. We then take the integral of $\LL \psi
|\sigma_1|\psi\RR$ over all possible input states $|\psi\RR$ and
we get the fidelity of 2/3.

\begin{figure}
\begin{center}
{\includegraphics[scale=0.8]{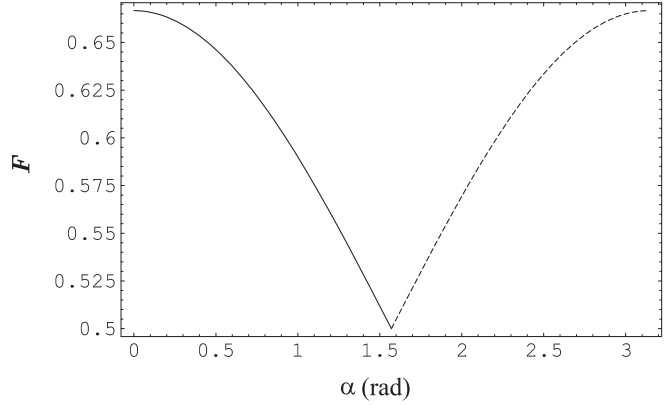}}
\end{center}
\caption{This graph shows the fidelity of measurement based
transformation and the phase angle $\alpha$.  The full line
represents the preparation $\sigma_1$ and the dotted line is for
$\sigma_2$. This result is identical to the usual measurement
based schemes considered in \cite{song1}. }
\label{Previous}\end{figure}

In \cite{buzek1,buzek2,gisin}, the U-NOT transformation has been
discussed. A measurement-based U-NOT operation has been discussed
in \cite{buzek2} which can be achieved as follows: If we get
positive with a guessed state, we prepare the orthogonal state to
the guessed state and when the result is negative we prepare the
guessed state, i.e.  \be \rho_2 = |\LL \psi|\phi\RR |^2
|\phi^{\perp}\RR\LL \phi^{\perp}| + |\LL \psi|\phi^{\perp}\RR|^2
|\phi\RR\LL \phi| \label{Rho3}\ee The result obtained was same as
the measurement fidelity of 2/3. We can also use a similar method
to that in (\ref{Rho2}).  We measure onto the $\{ |0\RR,|1\RR \}$
basis and when we get $|0\RR$ we prepare $|1\RR$ and for $|1\RR$,
$|0\RR$ is prepared.  This gives \be \sigma_2 = |\LL \psi|0\RR |^2
|1\RR\LL 1| + |\LL \psi |1\RR |^2 |0\RR\LL 0| \label{Rho4}\ee
After taking an integral of $\LL \psi |\sigma_2 |\psi\RR$ over all
$|\psi\RR$, we again obtain the same fidelity 2/3 as in the case
of $\rho_2$. In \cite{song1}, we considered a general linear
transformation \be (\vartheta,\varphi) \rightarrow (\vartheta -
\alpha,\varphi) \label{tra}\ee  Even in the case of transformation
(\ref{tra}), two density matrices $\sigma_1$ and $\sigma_2$ yield
the same results as obtained in \cite{song1}.  For $0\leq \alpha
\leq \pi/2$, $\sigma_1$ in (\ref{Rho2}) and for $\pi/2 \leq \alpha
\leq \pi$, $\sigma_2$ give the fidelity as shown in figure
\ref{Previous}.  This fidelity is same as the ones obtained by
$\rho_1$ in (\ref{Rho1}) and $\rho_2$ in (\ref{Rho3}).

We have shown that even with a specific basis $\{ |0\RR,|1\RR \}$,
we obtain the same results for measurement of arbitrary states,
the U-NOT and the general linear transformation cases. However we
will show in the following sections, this particular choice of
basis does not give the same results for other measurement based
nonlinear transformations.

\section{Nonlinear Rotations}
\begin{figure}
\begin{center}
{\includegraphics[scale=0.8]{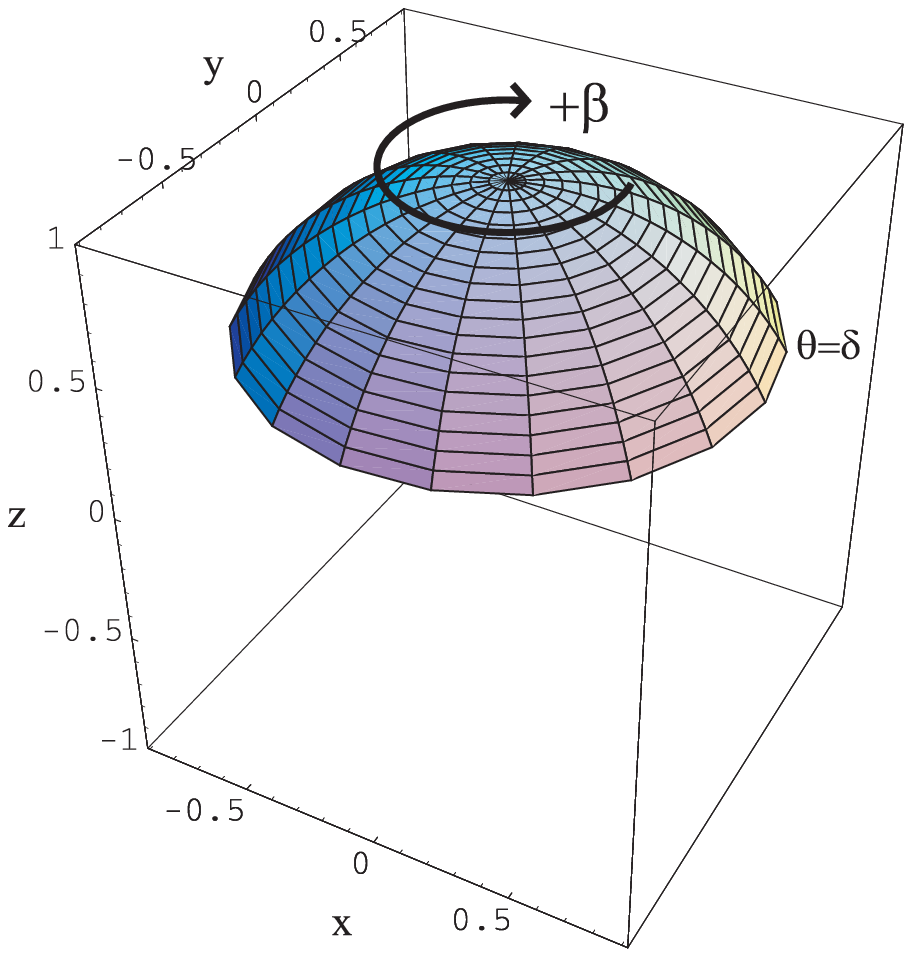}} $\;\;\; $
{\includegraphics[scale=0.8]{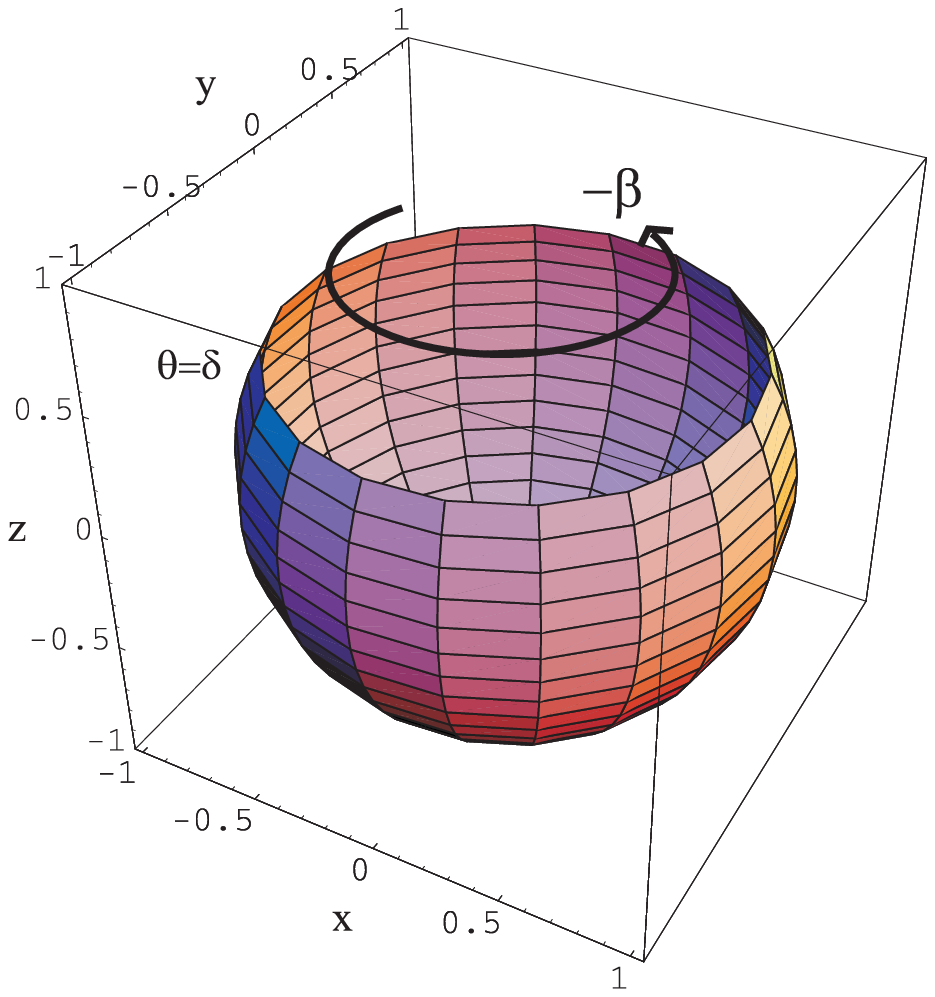}}
\end{center}
\caption{This figure shows nonlinear rotation of a Bloch sphere of
an unknown qubit.  For $\vartheta >\delta$, we rotate by $+\beta$
while for $\vartheta \leq \delta$ we rotate the qubit by $-\beta$
where $0\leq \delta \leq \pi$. } \label{rotationPI2}\end{figure}

It is well known that rotations of a qubit about the $z$-axis by
the angle $\beta$ can be achieved unitarily with the following
operator, \be R_z(\beta) = \left(
\begin{array}{cc} e^{-i\beta/2} & 0 \\ 0 & e^{i\beta/2}
\end{array} \right) \ee
Gisin has proposed \cite{gisin1} a rather different rotation of a
qubit. For some inputs $|\psi(\vartheta,\varphi)\RR$ we rotate by
the angle $\beta$ about the $z$-axis and for different
$(\vartheta,\varphi)$ we rotate by $-\beta$. Suppose given an
unknown state, we want to transform it as follows \be \left\{
\begin{array}{ll}
0\leq \vartheta < \delta, &  |\psi(\vartheta,\varphi)\RR \rightarrow |\psi(\vartheta,\varphi +\beta)\RR \\
\delta \leq \vartheta \leq \pi , &  |\psi(\vartheta,\varphi)\RR\rightarrow |\psi(\vartheta,\varphi - \beta)\RR \\
\end{array} \right. \label{tranROT}\ee  In a Bloch sphere, this can be drawn
as shown in figure \ref{rotationPI2}.
We consider two different approaches in achieving the
transformations in (\ref{tranROT}), measurement and quantum based
schemes.  We first consider a measurement scheme.

$   $
\newline
{\bf {Measurement Scheme}} $ \;\;\; $  Let us consider the case
when $\delta=\pi/2$. The first thing we could try is to use a
similar method to that in (\ref{Rho1}). That is, we measure onto
the guessed basis $\{ |\phi\RR,|\phi^{\perp}\RR\}$ (where
$|\phi\RR$ is chosen to be in the upper hemisphere) and rotate by
$+\beta$ if the result is positive and when $|\phi^{\perp}\RR$ is
obtained we rotate by $-\beta$. In this case, we have the density
matrix, \be \rho_3 = |\LL \psi|\phi\RR |^2 |\phi(\mu, \nu+\beta
)\RR\LL \phi (\mu, \nu+\beta )|+ |\LL\psi|\phi^{\perp}\RR |^2
|\phi^{\perp}(\mu, \nu-\beta ) \RR\LL \phi^{\perp} (\mu,\nu-\beta
)| \ee Then we need to take the average of $\rho_3$ for all
possible $|\phi(\mu,\nu)\RR$'s where $0\leq \mu \leq \pi/2$ and
$0\leq \nu \leq 2\pi$. If we take an example where $\beta=\pi/3$,
then the fidelity is obtained as \bea F &=&
\frac{1}{4\pi}\int_0^{2\pi} \int_0^{\delta=\pi/2} \LL \psi
(\vartheta,\varphi +\pi/3)| \overline{\rho_3}|
\psi(\vartheta,\varphi+\pi/3)\RR \sin\vartheta
\, d\vartheta d\varphi \nonumber \\
 &+& \frac{1}{4\pi} \int_0^{2\pi} \int_{\delta=\pi/2}^{\pi} \LL \psi
(\vartheta,\varphi-\pi/3)| \overline{\rho_3}|
\psi(\vartheta,\varphi-\pi/3)\RR \sin\vartheta \, d\vartheta
d\varphi = .58333... \label{F1}\eea  where $\overline{\rho_3}$ is
the averaged $\rho_3$. Can we do any better? Let us consider
another preparation, $\rho_1$ in (\ref{Rho1}), i.e. we do not
rotate by $+\beta$ or $-\beta$. Using the equation in (\ref{F1}),
instead of $\overline{\rho_3}$ we put $\overline{\rho_1}$, then
the fidelity is obtained as $.6111...$.

Let us consider another method that we considered in sect. 2, i.e.
with $\sigma_1$ in (\ref{Rho2}).  The fidelity is obtained as \bea
F &=& \frac{1}{4\pi}\int_0^{2\pi} \int_0^{\delta=\pi/2} \LL \psi
(\vartheta,\varphi
+\beta)|\sigma_1|\psi(\vartheta,\varphi+\beta)\RR \sin\vartheta d\vartheta d\varphi \nonumber \\
&+& \frac{1}{4\pi}\int_0^{2\pi} \int_{\delta=\pi/2}^{\pi} \LL
\psi(\vartheta,\varphi -\beta) |\sigma_1
|\psi(\vartheta,\varphi-\beta)\RR \sin\vartheta d\vartheta
d\varphi = \frac{2}{3} \label{Fvarrho}\eea This result is also
independent of the values $\beta$'s since the phase factor is
cancelled in $\LL \psi
(\vartheta,\varphi+\beta)|\sigma_1|\psi(\vartheta,\varphi+\beta)\RR$
and $\LL \psi(\vartheta,\varphi -\beta) |\sigma_1
|\psi(\vartheta,\varphi-\beta)\RR $. Also note that 2/3 can be
obtained for any $\delta$ as well. Since the quantity we are
integrating is independent of $\beta$, therefore $\LL
\psi(\vartheta,\varphi+\beta)|\sigma_1
|\psi(\vartheta,\varphi+\beta)\RR$ = $\LL
\psi(\vartheta,\varphi-\beta)|\sigma_1
|\psi(\vartheta,\varphi-\beta)\RR$.  Then we can put the two
integrals in (\ref{Fvarrho}) together which would give the same
result regardless of $\delta$ in a single integral. Therefore for
any $\beta$ or $\delta$, we can obtain the fidelity of 2/3
according to the transformation in (\ref{tranROT}) with the
measurement preparation of $\sigma_1$ of (\ref{Rho2}).

$    $
\newline
{\bf Quantum Scheme} $\;\;\; $ We now discuss how we can achieve
the transformation in (\ref{tranROT}) by unitary operations. One
can see that when $\delta=0,\pi$ then it can be achieved perfectly
by the unitary rotation about the $z$-axis. Let us consider the
most general transformation on a single qubit.  We consider a
unitary evolution on the single qubit and some ancilla prepared in
a known state $|Q\RR$ (this is taken to be normalised). This then
gives, \be  \begin{array}{l}
|0\RR|Q\RR \rightarrow  |1\RR|A\RR + |0\RR|B\RR  \\
|1\RR|Q\RR \rightarrow  |0\RR|\tA \RR + |1\RR|\tB\RR \end{array}
\label{Tran}\ee where $|A\RR,|\tA\RR,|B\RR,|\tB\RR$ may not be
normalised nor orthogonal to each other.  From orthogonality and
normalisation conditions of (\ref{Tran}), we have \bea
 &  & \begin{array}{l} |A|^2 +
|B|^2 = 1 \\ |\tA |^2 + |\tB|^2 = 1 \end{array} \label{condi1} \\
&  & \LL B|\tA\RR + \LL A|\tB \RR = 0 \label{condi2}\eea We
consider an unknown state in a Bloch sphere as in (\ref{bloch}) to
go through the transformation in (\ref{Tran}) and tracing over the
ancilla states yields the following density matrix, \bea
\rho^{out}& = & |0\RR\LL 1| ( \cos^2\frac{\vartheta}{2} \LL A|B\RR + \sin^2\frac{\vartheta}{2}\LL \tB|\tA \RR + \cos\frac{\vartheta}{2}\sin\frac{\vartheta}{2}e^{i\varphi} \LL A|\tA\RR + e^{-i\varphi}\cos\frac{\vartheta}{2} \sin\frac{\vartheta}{2} \LL \tB|B\RR ) \nonumber \\
                 & + & |1\RR\LL 0| ( \cos^2\frac{\vartheta}{2} \LL B|A\RR + \sin^2\frac{\vartheta}{2}\LL \tA|\tB\RR + \cos\frac{\vartheta}{2}\sin\frac{\vartheta}{2}e^{i\varphi} \LL B|\tB\RR + e^{-i\varphi}\cos\frac{\vartheta}{2} \sin\frac{\vartheta}{2} \LL \tA|A\RR
                 )\nonumber  \\
                  & + &|0\RR\LL 0| ( \cos^2\frac{\vartheta}{2} |B|^2 + \sin^2\frac{\vartheta}{2}|\tA|^2 + \cos\frac{\vartheta}{2}\sin\frac{\vartheta}{2}e^{i\varphi} \LL B|\tA\RR + e^{-i\varphi}\cos\frac{\vartheta}{2} \sin\frac{\vartheta}{2} \LL \tA|B\RR ) \nonumber \\
                 & + &|1\RR\LL 1| ( \cos^2\frac{\vartheta}{2} |A|^2 + \sin^2\frac{\vartheta}{2}|\tB|^2 + \cos\frac{\vartheta}{2}\sin\frac{\vartheta}{2}e^{i\varphi} \LL A|\tB\RR + e^{-i\varphi}\cos\frac{\vartheta}{2} \sin\frac{\vartheta}{2} \LL \tB|A\RR ) \label{RHOout}
\eea  Since we are considering a transformation which varies
depending on the value of the input, we need to get the average
value.  That is, the fidelity can be obtained by averaging over
$\vartheta$ and $\varphi$ instead of imposing a constraint such
that it is independent of $\vartheta$ and $\varphi$ as in the
universal transformation case.

\begin{figure}
\begin{center}
{\includegraphics[scale=.8]{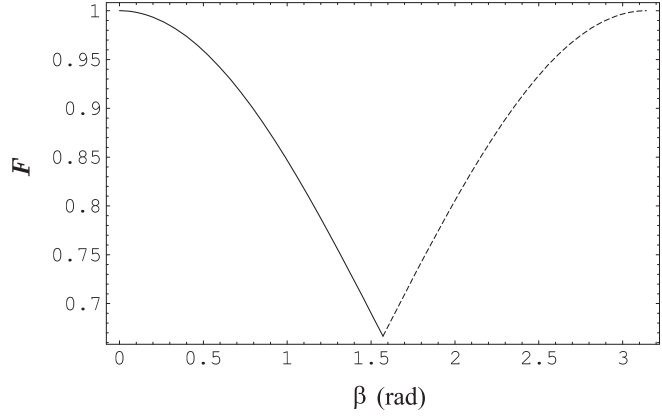}}
\end{center}
\caption{This is a graph for unitary transformation of nonlinear
rotation when $\delta=\pi/2$ for arbitrary $\beta$.  For $0\leq
\beta \leq \pi/2$, identity map yields the optimal map which is
shown as full line.  The dotted line corresponds to the rotation
by $\pi$ for $\pi/2 \leq \beta \leq \pi$. }
\label{Qrotpi}\end{figure}

Let us first consider the case when
$\delta=\pi/2$. The average fidelity is then \bea F &=&
\frac{1}{4\pi}\int_0^{2\pi} \int_0^{\pi/2} \LL \psi
(\vartheta,\varphi+\beta) |\rho^{out} |\psi
(\vartheta,\varphi+\beta) \RR \sin\vartheta d\vartheta d\varphi
\nonumber
\\ &+& \frac{1}{4\pi}\int_0^{2\pi} \int_{\pi/2}^{\pi} \LL \psi (\vartheta,\varphi-\beta)
|\rho^{out} |\psi(\vartheta,\varphi-\beta)\RR \sin\vartheta
d\vartheta d\varphi \eea  which comes out as \be F= \frac{1}{3} +
\frac{1}{6} (|B|^2 + |\tB|^2) +\frac{\cos \beta}{6}(\LL B|\tB \RR
+ \LL \tB |B\RR ) \label{fiD}\ee  We now would like to find the
values of $|B|,|\tB|,\LL B|\tB\RR$ and $\LL \tB|B\RR$ such that
the fidelity $F$ in (\ref{fiD}) is the highest. We know $\cos
\beta$ is positive for $0 < \beta < \pi/2$ while it's negative for
$\pi/2 < \beta < \pi$. Therefore when $0 \leq \beta \leq \pi/2$,
$|B|^2=1= |\tB|^2$ and $\LL B|\tB\RR = 1=\LL \tB|B\RR$ yields the
highest fidelity which can be satisfied with
$|B\RR=|0\RR$,$|\tB\RR=|0\RR$.  This implies that we have
$|A\RR=0=|\tA\RR$ from (\ref{condi1}). Similarly, for $\pi/2 \leq
\beta \leq \pi$, $|B|^2=1=|B|^2$ and $\LL B|\tB \RR = -1 =\LL
\tB|B\RR$ would give the optimal fidelity. This can be done with
the choice of $|B\RR=|0\RR$ and $|\tB\RR=-|0\RR$. This latter map
corresponds to the unitary rotation by $\pi$ about the $z$-axis
while the former map is simply an identity map. The fidelity is
given as \be \frac{2}{3} \leq F \leq 1 \ee where the actual
dependence on the value of $\beta$ is shown in figure
\ref{Qrotpi}.
  One can easily see that when $\beta=0$
and $\beta=\pi$, it should be achieved perfectly therefore the
fidelity is equal to 1. The fidelity has the lowest value of 2/3
when $\beta=\pi/2$.

\begin{figure}
\begin{center}
{\includegraphics[scale=1.1]{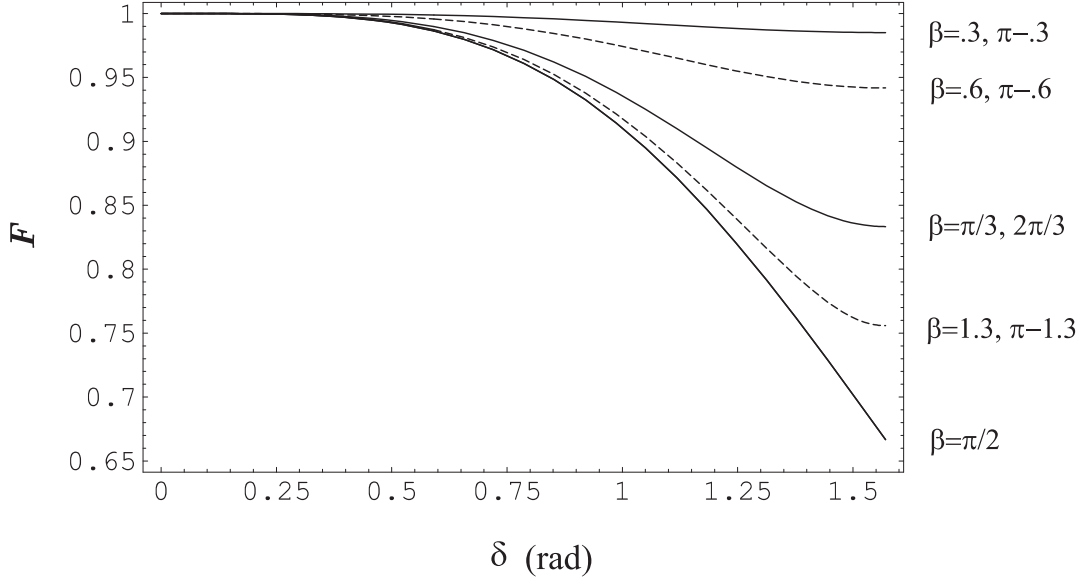}}
\end{center}
\caption{This graph shows the fidelities of quantum scheme for
nonlinear rotation as a function of $\delta$ for different values
of $\beta$.  The end points of each line, i.e. at $\delta=\pi/2$
corresponds to the graph in figure \ref{Qrotpi}.}
\label{b}\end{figure}

We now would like to consider a case for arbitrary $\delta$, i.e.
we want  \be \left\{
\begin{array}{ll}
0\leq \vartheta \leq \delta &  {\rm rotation \; by} +\beta \\
\delta \leq \vartheta \leq \pi &  {\rm rotation \; by} -\beta \\
\end{array} \right. \label{ARBbeta}\ee  We consider the same output density
matrix $\rho^{out}$ in (\ref{RHOout}) and the fidelity for the
transformation (\ref{ARBbeta}) is obtained as follows, \bea F &=&
\frac{1}{4\pi} \int_0^{2\pi} \int_0^{\delta} \LL \psi
(\vartheta,\varphi+\beta) |\rho^{out} |\psi
(\vartheta,\varphi+\beta) \RR \sin\vartheta d\vartheta d\varphi
\nonumber
\\ &+& \frac{1}{4\pi}\int_0^{2\pi} \int_{\delta}^{\pi} \LL \psi (\vartheta,\varphi-\beta)
|\rho^{out} |\psi(\vartheta,\varphi-\beta)\RR \sin\vartheta
d\vartheta d\varphi \label{F5}\eea Let us consider a specific
example when $\beta=\pi/3$. Then the fidelity in (\ref{F5}) comes
out as \bea F &=& \frac{1}{3} + \frac{1}{6} (|B|^2 + |\tB|^2)
+\left(.08333-(.1623 \cos\delta - .018 \cos3\delta)i\right)\LL
B|\tB \RR \nonumber \\ & & + \left(.08333+(.1623 \cos\delta - .018
\cos3\delta)i\right) \LL \tB |B\RR \label{F30}\eea We need to find
the values of $|B|^2$,$|\tB|^2$,$\LL B|\tB\RR$ and $\LL \tB|B\RR$
such that the fidelity $F$ yields the highest value. Suppose $ \LL
B|\tB\RR = a+bi$ and $\LL \tB|B\RR = a-bi $ where $a,b$ are real.
Then the two terms of $\LL B|\tB\RR$ and $\LL \tB|B\RR$ in
(\ref{F30}) are written as \be 2(.08333) a + 2(.1623 \cos\delta -
.018 \cos3\delta) b \label{TWO}\ee  One can verify that $(.1623
\cos\delta - .018 \cos3\delta)$ is non-negative for $0\leq \delta
\leq \pi/2$, hence we know that these two terms will be maximum
with an appropriate $a$ and $b$ when $a^2+b^2=1$ rather than
$a^2+b^2 <1$. Therefore we can choose $|B\RR = e^{-i\chi/2}|0\RR$
and $|\tB\RR = e^{i\chi/2}|0\RR$ which implies $\LL B|\tB\RR =
e^{i\chi}$ ($\LL \tB|B\RR = e^{-i\chi}$). We can obtain $\chi$
which would give the $F$ maximum value as follows, \be
\chi=\arccos \left( \frac{83333. \times 10^{-3} }{\sqrt{6.94
+2.63\times 10\cos^2 \delta -5.85 \cos\delta \cos 3\delta +
3.25\times 10^{-1} \cos^2 3\delta }} \right)\label{w}\ee
\begin{figure}
\begin{center}
{\includegraphics[scale=.8]{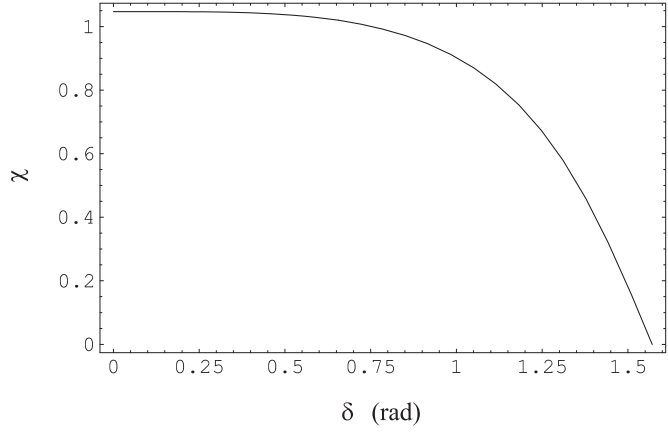}}
\end{center}
\caption{This is a graph between $\chi$ in (\ref{w}) and $\delta$.
This value of $\chi$ for a given $\delta$ gives the maximum
fidelity in (\ref{F5}) } \label{a}\end{figure} Therefore $\chi$ is
the angle that you want to rotate by and the graph of the value
$\chi$ and the angle $\delta$ is shown in figure \ref{a}. It shows
that when $\delta=0$ it is best to rotate by $\pi/3$ and for
$\delta=\pi/2$ identity map is optimal as we have checked
previously. By symmetry, the case of $\pi/2 \leq \delta \leq \pi$
can be obtained similarly except that we rotate by opposite
direction. Other cases of $\beta$ can be achieved using the
similar methods described above. In figure \ref{b}, we show the
graph of fidelity and $\delta$ for different $\beta$'s.

Therefore quantum scheme is always better than or equal to the
measurement scheme (which is $F=2/3$) as expected.

\section{Nonlinear ORTHOG Transformations}
\begin{figure}
\begin{center}
{\includegraphics[scale=0.8]{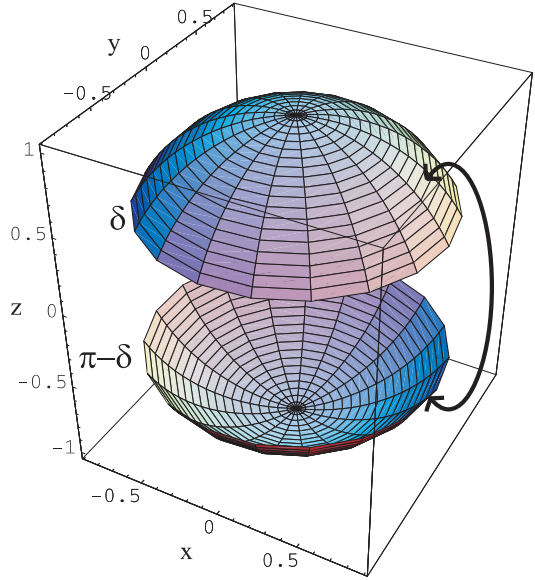}}
\end{center}
\caption{We want to apply ORTHOG operation only to the shown areas
of a Bloch sphere, i.e. $0\leq \vartheta \leq \delta$ and
$\pi-\delta \leq \vartheta \leq \pi$ while leaving other areas of
a Bloch sphere unchanged. } \label{Unot}\end{figure}

A NOT-gate is one of the simple operations performed on a
classical bit, 0 to 1 and 1 to 0.  Just as in the cloning case,
the operation which sends a qubit to its orthogonal state is not
so simple. In \cite{buzek1,buzek2,gisin1}, U-NOT operation for
qubits has been studied. Since a perfect U-NOT gate would require
anti-unitary operation, they considered the optimal way of
imitating perfect U-NOT transformation and showed the optimal
fidelity to be 2/3 for a single qubit input.  This fidelity was
same as the measurement fidelity which we reviewed in sect. 2.
Therefore measuring the qubit then preparing another qubit
orthogonal to the original one and the quantum unitary
transformation were shown to have equal efficiency.  In this
section, we generalise this result and consider nonlinear ORTHOG
transformations. As shown in figure \ref{Unot}, we want to apply
ORTHOG-gate to the shown area of a Bloch sphere, i.e. \be \left\{
\begin{array}{ll}
0\leq \vartheta \leq \delta, \; \pi-\delta \leq \vartheta \leq \pi  & \;\; \;\;\; |\psi(\vartheta,\varphi)\RR \rightarrow |\psi(\vartheta-\pi,\varphi)\RR  \\
\delta \leq \vartheta \leq \pi - \delta  & \;\;\;\; \; |\psi(\vartheta,\varphi)\RR\rightarrow |\psi(\vartheta,\varphi)\RR \\
\end{array} \right. \label{nonUNOT}\ee
 When
$\delta=0$, this transformation is simply an identity map and when
$\delta=\pi/2$, it would correspond to the U-NOT gate (if a
universality condition is imposed).

$     $
\newline
{\bf Measurement Scheme} $\;\;\; $ In the previous sections, we
considered two types of measurement schemes.  We now consider
these two types in achieving the transformation in
(\ref{nonUNOT}).  For the first type, we could consider the
density matrices $\rho_1$ in (\ref{Rho1}) and $\rho_2$ in
(\ref{Rho3}).  We take the average over all $|\phi\RR$'s then the
fidelity is obtained by \be F_{1,2} = \int_0^{2\pi}
\int_0^{\delta} \LL \psi^{\perp} |\overline{\rho_{1,2}}
|\psi^{\perp} \RR d\Omega +
\int_0^{2\pi}\int_{\delta}^{\pi-\delta} \LL \psi
|\overline{\rho_{1,2}} |\psi\RR d\Omega + \int_0^{2\pi}
\int_{\pi-\delta}^{\pi} \LL
\psi^{\perp}|\overline{\rho_{1,2}}|\psi^{\perp}\RR d\Omega \ee
where $d\Omega = 1/(4\pi) \sin\vartheta d\vartheta d\varphi$.
Another way is to prepare $\sigma_1$ in (\ref{Rho2}) and
$\sigma_2$ in (\ref{Rho4}).
\begin{figure}
\begin{center}
{\includegraphics[scale=1]{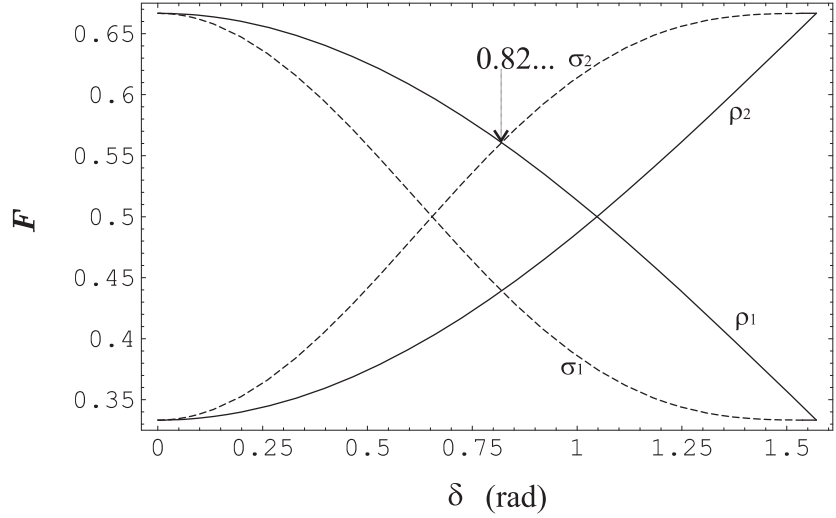}}
\end{center}
\caption{$\rho_1$ and $\rho_2$ are represented by full lines and
the dotted lines represent $\sigma_1$ and $\sigma_2$. For $0\leq
\delta \leq .82...$, $\rho_1$ is the optimal preparation while for
$.82... \leq \delta \leq \pi/2$, it is the $\sigma_2$ which gives
the higher fidelity. } \label{future337}\end{figure} Although the
measurement fidelity, the measurement based U-NOT and the
measurement based general linear transformation all yielded the
same results for both preparations of $\rho$ and $\sigma$ as shown
in sect. 2, it is no longer true in case of nonlinear ORTHOG
transformations. The fidelities for $\rho_{1,2},\sigma_{1,2}$ are
obtained as shown in figure \ref{future337}. Interestingly, for
$0\leq \delta \leq .82...$ $\rho_1$ is the better preparation
while for $.82... \leq \delta \leq \pi/2$, it is the $\sigma_2$
preparation which gives the higher fidelity.

$    $
\newline
{\bf Quantum Scheme} $\;\;\; $  We consider the same
transformation (\ref{Tran}) and the same conditions
(\ref{condi1},\ref{condi2}). With $\rho^{out}$ in (\ref{RHOout}),
the fidelity is \be F = \int_0^{2\pi} \int_0^{\delta} \LL
\psi^{\perp} |\rho^{out} |\psi^{\perp}\RR d\Omega + \int_0^{2\pi}
\int_{\delta}^{\pi -\delta} \LL \psi |\rho^{out} |\psi\RR d\Omega
+ \int_0^{2\pi} \int_{\pi-\delta}^{\pi} \LL \psi^{\perp}
|\rho^{out} |\psi^{\perp}\RR d\Omega  \label{F8}\ee where $d\Omega
= 1/(4\pi) \sin\vartheta d\vartheta d\varphi$. Let us consider the
simplest case when $\delta = \pi/2$ which corresponds to the
ORTHOG gate. In that case, the fidelity in (\ref{F8}) becomes \bea
F &=& \frac{1}{3} |A|^2 + \frac{1}{3} |\tA |^2 + \frac{1}{6}
(|B|^2 + |\tB|^2 - \LL B|\tB\RR
- \LL \tB |B\RR ) \\
  &=& \frac{2}{3} -\frac{1}{6} (|B|^2 + |\tB|^2 + \LL B|\tB\RR
+ \LL \tB |B\RR ) \label{F9}\eea  Suppose $|B\RR=(a+bi)|0\RR$ and
$|\tB\RR=(c+di)|0\RR$ where $a,b,c,d$ are real. This implies \be
|B|^2 = a^2 + b^2 \;, \; |\tB|^2 = c^2 + d^2 \ee then \be |B|^2 +
|\tB|^2 + \LL B|\tB \RR + \LL \tB |B\RR = (a+c)^2 +(b+d)^2
\label{newC1}\ee This quantity is always non-negative and the
minimum is achieved with $a=-c$ and $b=-d$ therefore \be |B\RR =
-|\tB\RR \; , \; |A|^2 = |\tA|^2 \label{choic1}\ee  What if
$|B\RR$ and $|\tB\RR$ are not the vectors pointing in the same
direction? Since $|B|^2 + |\tB|^2 + \LL B|\tB\RR + \LL \tB |B\RR$
= $| |B\RR + |\tB\RR |^2$ $\geq 0$, we know (\ref{choic1}) is
optimal. The U-NOT transformation derived in \cite{buzek1,buzek2}
is \be |A\RR = -\sqrt{\frac{2}{3}} |00\RR,\, |\tA\RR =
\sqrt{\frac{2}{3}} |11\RR,\,  |B\RR = \sqrt{\frac{1}{6}} (|01\RR +
|10\RR ),\, |\tB\RR = -\sqrt{\frac{1}{6}} (|01\RR + |10\RR )
\label{UnotBuzek}\ee  This is one special case of (\ref{choic1})
which is consistent considering the universality of
(\ref{UnotBuzek}).

\begin{figure}
\begin{center}
{\includegraphics[scale=.8]{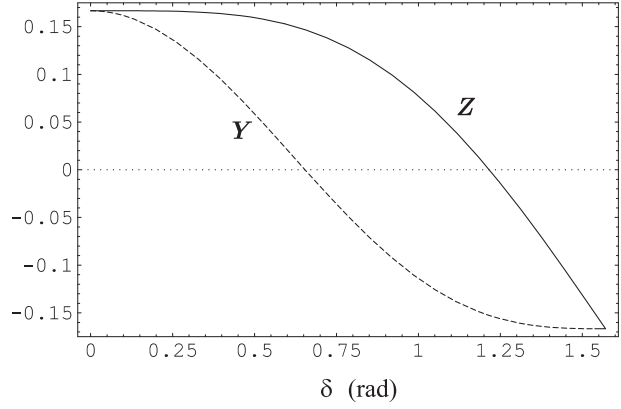}}
\end{center}
\caption{This shows the coefficients $Y$ and $Z$ in (\ref{yy}) and
(\ref{zz}) as a function of $\delta$. }
\label{coeffYZ}\end{figure}

For arbitrary $\delta$, the fidelity is obtained using the formula
in (\ref{F8}) which is given as \be F = X + Y(|B|^2+|\tB|^2 ) +
Z(\LL B|\tB\RR + \LL \tB |B\RR ) \label{F10}\ee where \bea X &
\equiv & \frac{2}{3} + \frac{1}{24}(\cos3(\pi-\delta)-\cos3\delta)
- \frac{1}{4} \cos\delta \\ Y & \equiv &  -\frac{1}{6} -
\frac{1}{8} (\cos2(\pi-\delta)-\cos2\delta ) -
\frac{1}{24}(\cos3(\pi-\delta)-\cos3\delta ) + \frac{1}{4}
\cos\delta \label{yy}\\ Z & \equiv & -\frac{1}{6} +
\frac{1}{48}(\cos3(\pi-\delta)-\cos3\delta) + \frac{3}{8}
\cos\delta \label{zz}\eea  We would like to find the values of
$|B|^2,|\tB|^2,\LL B|\tB\RR$ and $\LL \tB|B\RR$ such that the
fidelity in (\ref{F10}) is maximum. For $0 \leq \delta \leq
\pi/2$, $X$ is always positive.  The values of $Y$ and $Z$ as
functions of $\delta$ are shown in figure \ref{coeffYZ}. When
$Y\geq 0$ and $Z\geq 0$, then the optimal map can be obtained when
$|B\RR=|0\RR=|\tB\RR$ which is simply an identity map.

Let us consider the case when $Y<0$ and $Z>0$ in particular when
$|Y| \leq |Z|$. As before, suppose $|B\RR=(a+bi)|0\RR$ and
$|\tB\RR=(c+di)|0\RR$ where $a,b,c,d$ are real numbers.  Then \bea
& &
Y(B^2+\tB^2)+Z(\LL B|\tB\RR + \LL \tB|B\RR ) \nonumber \\
&=& Y(a^2+b^2+c^2+d^2)+ Z(2ac+2bd) \label{secondLINE} \\
&=& Y\left( (a-c)^2 + (b-d)^2)\right) +
(Z-Y)(2ac+2bd)\label{thirdLINE} \eea  Since $Y\leq 0$, $a=c$ and
$b=d$ will give the $Y$ term in (\ref{thirdLINE}) highest, which
is zero. Let us consider the term $(ac +bd)$.  Since we are
assuming $|B\RR$ and $|\tB\RR$ may not be normalised, we know
$a^2+b^2\leq 1$ and $c^2+d^2\leq 1$. We know that the highest
value for $ac+bd$ will be attained when $a^2+b^2=1$ and
$c^2+d^2=1$. Let us consider
\bea & & ac + bd \stackrel{?}{\leq}  1  \label{1line}\\ & \Rightarrow & ac + \sqrt{(1-a^2)(1-c^2)} \leq  1  \label{2line}\\
& \Rightarrow & (1-a^2)(1-c^2) \leq (1-ac)^2  \label{3line}\\  &
\Rightarrow & 0 \leq a^2-2ac+c^2 = (a-c)^2 \label{4line}\eea
Therefore from (\ref{4line}), we know (\ref{1line}) is true.
Therefore the highest possible value for the $(Z-Y)$ term is
achieved when $a=c$ (which we choose to be 1) and this also gives
the lowest possible value for the $Y$ term.  If $B$ and $\tB$ were
the vectors pointing different directions, the second term in
(\ref{secondLINE}) which is positive will be even less.  Therefore
our choice of the same state $|0\RR$ (or any other normalised
state) for $|B\RR$ and $|\tB\RR$ is indeed optimal.

Let us consider the case when $|Y| \geq |Z|$ for $Y<0$ and $Z>0$.
Since $a^2+c^2\geq \pm 2ac$ and $b^2+d^2 \geq \pm 2bd$, we know
(\ref{secondLINE}) is non-positive.  Therefore the choice of
$|B\RR=0=|\tB\RR$ yields the highest fidelity. For a region where
$Y<0$ and $Z<0$, one can see $|Y|>|Z|$ from figure \ref{coeffYZ}.
Therefore again $|B\RR=0=|\tB\RR$ is the optimal map. This map
corresponds to the transformation $|0\RR \rightarrow |1\RR$,
$|1\RR \rightarrow |0\RR$ which is one of the
Pauli matrices $\sigma_x\equiv \left( \begin{array}{cc}  0  & 1 \\
1 & 0
\end{array} \right)$. The final optimal map is obtained as follows, \be \left\{
\begin{array}{ll}
0\leq \delta \leq .932197... & \;\;\;\;\;\;    {\bf 1} \\
.932197...  \leq \delta \leq \pi/2  &  \;\;\;\;\;\;  |0\RR \rightarrow |1\RR,\, |1\RR \rightarrow |0\RR  \\
\end{array} \right. \label{NonUNOT}\ee
where $\delta = .932197...$ corresponds to the point when
$|Y|=|Z|$ for the period $Y<0$ and $Z>0$.  Therefore the optimal
map is either an identity map or a $\sigma_x$ unitary operation
depending on the value of $\delta$.

\begin{figure}
\begin{center}
{\includegraphics[scale=.9]{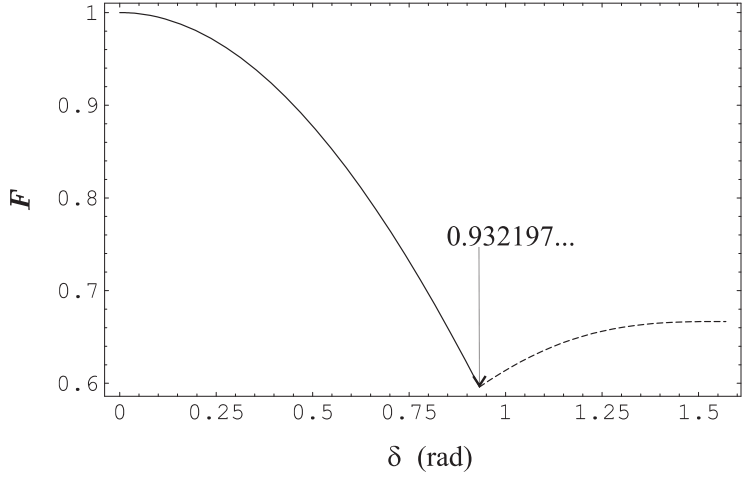}}
\end{center}
\caption{This is a graph for nonlinear ORTHOG transformation.  It
shows the fidelity $F$ as a function of $\delta$ where the full
line is an identity map and the dotted line represents the unitary
operation of Pauli matrix $\sigma_x$.}
\label{future338}\end{figure}

\section{Nonlinear General Transformations}
\begin{figure}
\begin{center}
{\includegraphics[scale=.8]{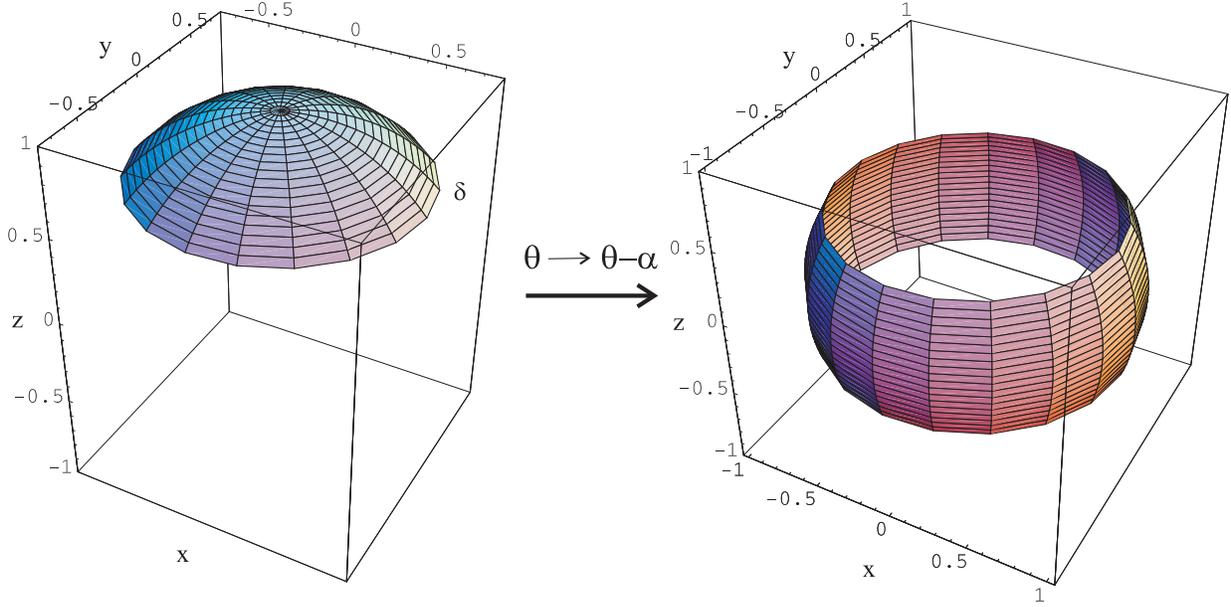}}
\end{center}
\caption{This figure shows the nonlinear general transformation.
We want to transform the shown area of a Bloch sphere by
$(\vartheta,\varphi)\rightarrow (\vartheta -\alpha,\varphi)$ while
leaving other areas of a Bloch sphere unchanged. }
\label{arbit}\end{figure}

Motivated by the U-NOT gate, we investigated a general arbitrary
manipulation that can be performed on a single qubit by studying
the universal linear transformation
$(\vartheta,\varphi)\rightarrow (\vartheta-\alpha,\varphi)$ in
\cite{song1}.  Rather surprisingly, we have found there are only
two maps which optimise such transformation, either identity or
U-NOT gate depending on $\alpha$. In this section, we attempt to
generalise this result to nonlinear cases. As shown in figure
\ref{arbit}, we want to transform $|\psi(\vartheta,\varphi)\RR$
into $|\psi(\vartheta-\alpha,\varphi)\RR$ only when $\vartheta$ is
between $0$ and an arbitrary angle $\delta$, i.e., \be \left\{
\begin{array}{ll}
0\leq \vartheta \leq \delta , &  |\psi(\vartheta,\varphi)\RR \rightarrow |\psi(\vartheta-\alpha,\varphi)\RR \\
\delta < \vartheta \leq \pi , &  |\psi(\vartheta,\varphi)\RR\rightarrow |\psi(\vartheta,\varphi)\RR \\
\end{array} \right. \label{TrA}\ee  When $\delta=\pi$, it would correspond
to what we have already considered for the universal
transformation in \cite{song1}.

\begin{figure}
\begin{center}
{\includegraphics[scale=.7]{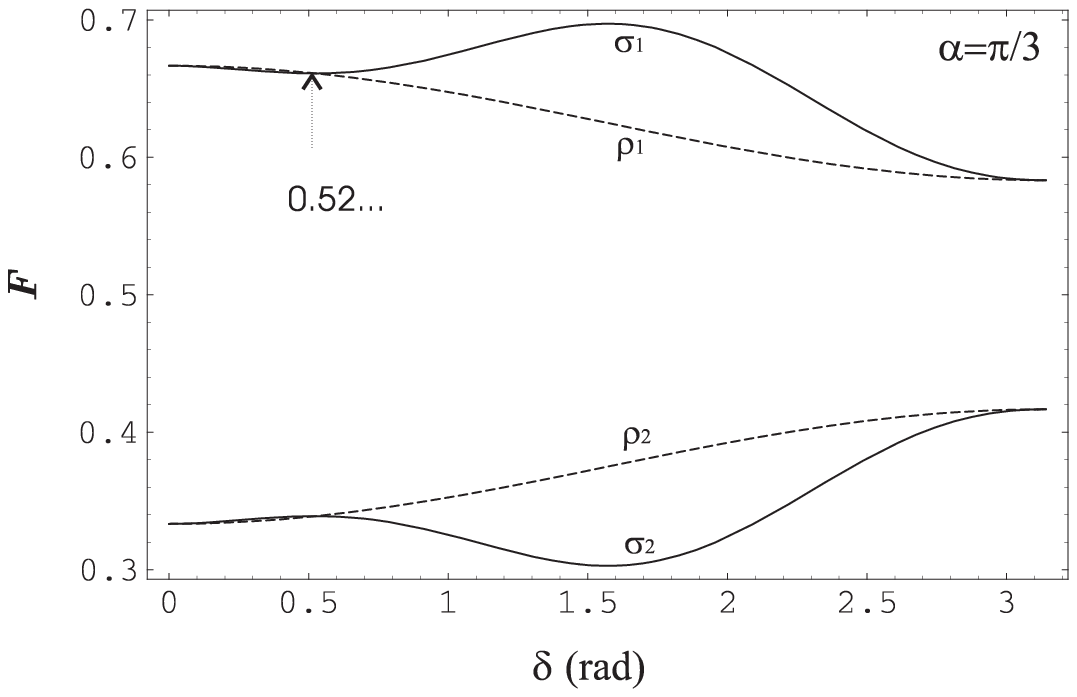}} $\;\;$
{\includegraphics[scale=.7]{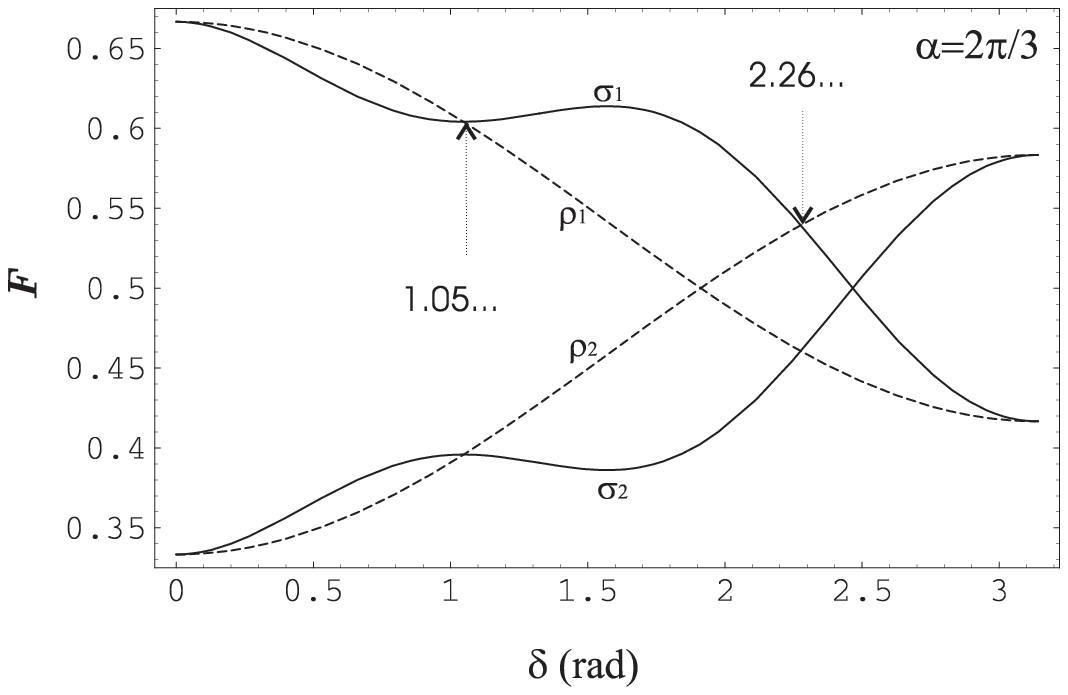}}
\end{center}
\caption{These two graphs show the fidelity of measurement scheme
for nonlinear general transformations. The left graph is for
$\alpha=\pi/3$ and the right one is when $\alpha=2\pi/3$. The full
lines are for $\sigma_{1,2}$ and the dotted lines represent
$\rho_{1,2}$. } \label{future339}\end{figure}

$    $
\newline
{\bf Measurement Scheme} $\;\;\; $ As in the nonlinear ORTHOG
transformation case, we consider the same density matrices
$\rho_{1,2}$ in (\ref{Rho1},\ref{Rho3}) and $\sigma_{1,2}$ in
(\ref{Rho2},\ref{Rho4}). We also take two examples of $\alpha$,
$\alpha=\pi/3$ and $\alpha=2\pi/3$. Then the fidelity is obtained
as follows, \be F_{1,2}= \int_0^{\delta} \LL
\psi(\vartheta-\alpha,\varphi)|\overline{\rho_{1,2}}|\psi(\vartheta-\alpha,\varphi)\RR
d\Omega + \int_{\delta}^{\pi} \LL
\psi(\vartheta,\varphi)|\overline{\rho_{1,2}}|\psi(\vartheta,\varphi)\RR
d\Omega  \ee where $0\leq \delta \leq \pi$ and similarly for
$\sigma_{1,2}$. As in the nonlinear ORTHOG case, these two
preparations yield different results that are shown in figure
\ref{future339}. Therefore the following maps have higher
fidelities at least among the measurement preparations,
$\rho_{1,2}$ and $\sigma_{1,2}$, \be \left\{
\begin{array}{ll}
0\leq \delta \leq 0.52... , &  \;\; \rho_1 \\
0.52... \leq \delta \leq \pi , &  \;\; \sigma_1
\end{array} \right. \ee for $\alpha=\pi/3$ and \be \left\{
\begin{array}{ll}
0\leq \delta \leq 1.05... , &  \;\; \rho_1 \\
1.05... \leq \delta \leq 2.26... , &  \;\; \sigma_1 \\
2.26... \leq \delta \leq \pi , & \;\; \rho_2
\end{array} \right. \ee when $\alpha=2\pi/3$.  Let us now
investigate a quantum case.

$     $
\newline
{\bf Quantum Scheme} $\;\;\; $ Let us first consider the case
which we showed in our previous paper \cite{song1}, i.e. the
transformation in (\ref{TrA}) and we take $\delta$ to be $\pi$.
Then the fidelity is obtained as follows,
\begin{figure}
\begin{center}
{\includegraphics[scale=.8]{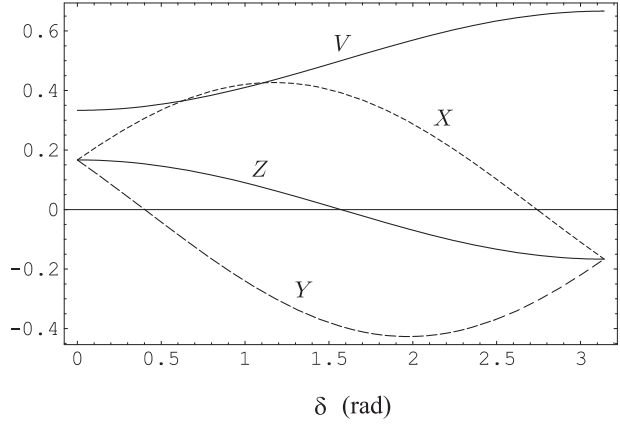}}
\end{center}
\caption{This figure shows the coefficients $V,X,Y,Z$ in
(\ref{xxx},\ref{zzz}) as a function of $\delta$.}
\label{pRev}\end{figure} \bea F&=& \int_0^{\delta =\pi} \LL
\psi(\vartheta-\alpha,\varphi)|\rho^{out}|\psi(\vartheta-\alpha,\varphi)\RR
d\Omega \\ &=& V + X |B|^2 + Y |\tB|^2 + Z(\LL B|\tB \RR + \LL \tB
|B\RR ) \label{var11} \eea where \bea  V&=& \frac{1}{2}
-\frac{1}{6}\cos\alpha ,  \;\;\; X = \frac{1}{6} \cos\alpha +.3926
\sin\alpha  \label{xxx}\\  Y &=& \frac{1}{6} \cos\alpha - .3926
\sin\alpha , \;\;\;\;\; Z = \frac{1}{6} \cos\alpha \label{zzz}
\eea  The values of $V,X,Y$ and $Z$ as functions of $\delta$ are
shown in figure \ref{pRev}. Since $V$ is always positive for all
$\alpha$, let us consider the other three terms. When $X,Y$ are
both positive, then the maximum $F$ is obtained when the
$|B|^2=|\tB|^2=\LL B|\tB\RR = \LL \tB|B\RR =1$ which can be
achieved easily with the choice of $|B\RR =|\tB\RR = |0\RR$, for
instance.

We consider the case when $X,Y$ are both negative. As before we
assume $|B\RR=(a+bi)|0\RR$ and $|\tB\RR=(c+di)|0\RR$. Then
(\ref{var11}) is rewritten as follows \be V + X(a^2 + b^2) + Y(c^2
+ d^2) + Z(2ac+2bd)\label{varied}\ee  Since the $Z$ term can
always be non-negative with an appropriate sign choices of
$a,b,c,d$, we do not need to worry about case of different vectors
(i.e. different directions) for $|B\RR$ and $|\tB\RR$ which would
only lessen the value of $Z$ term. The $X,Y$ and $Z$ terms in
(\ref{varied}) can be rewritten as follows, \bea &  & X a^2 + Yc^2 + Z(2ac) +Xb^2+Yd^2 + Z(2bd) \\
&=&Y(c+\frac{Z}{Y} a)^2 +Y(d+\frac{Z}{Y} b)^2+
(X-\frac{Z^2}{Y})(a^2+b^2) \label{varied3} \eea One can check that
$(X-Z^2/Y) >0 $ for the region where $X,Y<0$. Therefore with the
choice of $a=1,b=0$ and $c=-Z/Y,d=0$, the fidelity will be the
highest.

\begin{figure}
\begin{center}
{\includegraphics[scale=.9]{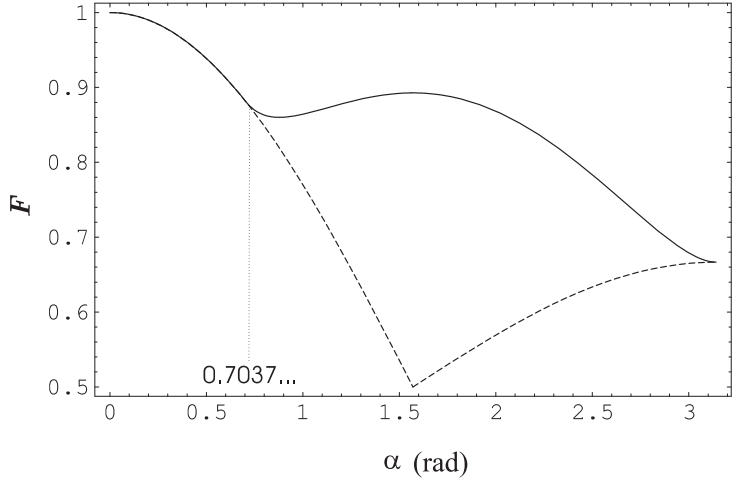}}
\end{center}
\caption{This figure shows the fidelities as a function of
$\alpha$ when $\delta=\pi$.  The dotted line is for universal
general linear transformation as shown in \cite{song1} and the
full line represents the averaged general linear transformation
which has substantially higher fidelity for $\alpha\geq .7037...$.
} \label{prev}\end{figure}

Let us consider the case when $X>0$ and $Y<0$. In this case, we
could consider (\ref{varied3}) again. The first two terms are
always non-positive and the second term is non-negative because
$X-Z^2/Y>0$ for the region where $X>0,Y<0$. Then $a=1$ and
$c=-Z/Y$ would yield the optimal fidelity. However $|Z/Y|>1$ when
$Y$ is near 0.  We then consider the equation in (\ref{varied})
again. Since $X$ is positive, we know $a^2+b^2=1$ would yield the
highest fidelity. Therefore we have (without the $V$ term), \be  X
+ Y(c^2+d^2)+Z(2ac+2\sqrt{1-a^2}d) \ee and we assume $a,c,d$ to be
non-negative. By taking partial derivatives, we can obtain the
maximum value. In this case, it is the end point, $c=1$ (and
$a=1$) that gives the optimal fidelity. The final graph can be
obtained as shown in figure \ref{prev}. Note that unlike the case
of nonlinear ORTHOG transformation (where the optimal map was
either identity or $\sigma_x$), the mapping is continuous for some
regions. It shows that the fidelity of averaged general
transformation (full line) is higher than the universal general
operation case (dotted line) that was obtained in \cite{song1} for
$\alpha\geq .7037...$.

\begin{figure}
\begin{center}
{\includegraphics[scale=1.2]{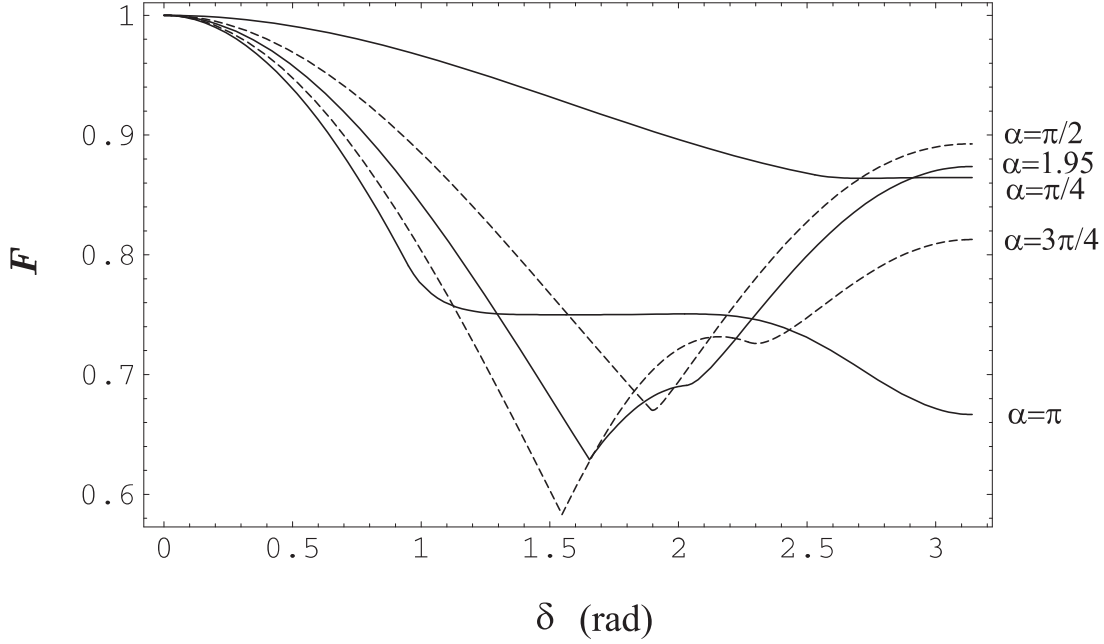}}
\end{center}
\caption{This is graph of nonlinear general transformations.  The
fidelities for different values of $\alpha$ as functions of
$\delta$ are shown. The end points of each line (i.e. when
$\delta=\pi$ corresponds to the full line in figure \ref{prev}. }
\label{qarbi}\end{figure}

Next we consider the proposed map in (\ref{TrA}) for arbitrary
$\delta$.  The fidelity is then \be F= \int_0^{\delta} \LL
\psi(\vartheta-\alpha,\varphi)|\rho^{out}|\psi(\vartheta-\alpha,\varphi)\RR
d\Omega + \int_{\delta}^{\pi} \LL
\psi(\vartheta,\varphi)|\rho^{out}|\psi(\vartheta,\varphi)\RR
d\Omega  \ee  Using a similar method as in the case of
$\delta=\pi$ we considered above, we can obtain the optimal
fidelities as shown in figure \ref{qarbi}.  In the figure, we show
five different values of $\alpha$ where the end point of each line
(i.e. when $\delta=\pi$) corresponds to the full line we obtained
in figure \ref{prev}.

\section{Remarks}
In this paper, we have studied three types of nonlinear
operations.  Firstly, we considered nonlinear rotations that
rotate different parts of a Bloch sphere (i.e. $0\leq \vartheta
\leq \delta$ and $\delta \leq \vartheta \leq \pi$) in opposite
directions about the $z$-axis. In case of a measurement-based
scheme, we showed that a measurement onto the particular (rather
than arbitrary) basis, $\{ |0\RR,|1\RR \}$, gives 2/3 fidelity.
For a quantum unitary transformation, we considered average
rotation manipulations. We showed that when the rotations are
applied to upper and lower hemispheres of a Bloch sphere (i.e.
when $\delta=\pi/2$), the optimal map is either identity (for
$0\leq \beta \leq \pi/2$) or the unitary rotation by $\pi$ (for
$\pi/2 \leq \beta \leq \pi$). For general $\delta$'s, we found the
optimal maps for different values of the rotating angle $\beta$.
In these cases, the optimal operations correspond to unitary
rotations by $\chi(\delta)$ about the $z$-axis.

Secondly, we studied nonlinear ORTHOG transformations where the
operations were applied only to $0\leq \vartheta \leq \delta$ and
$\pi-\delta \leq \vartheta \leq \pi$.  For a measurement scheme,
we found the preparations $\rho$'s and $\sigma$'s yield different
results.  For $0\leq \delta \leq .82...$, $\rho_1$ in (\ref{Rho1})
yields a higher fidelity while for $.82... \leq \delta \leq
\pi/2$, $\sigma_2$ in (\ref{Rho4}) is a better preparation. In
case of a quantum scheme, we considered a special case when the
ORTHOG operation is applied to a whole Bloch sphere (i.e. when
$\delta = \pi/2$).  In this case, we found Bu\v{z}ek et al.'s
U-NOT gate appears as a special case of average ORTHOG
transformation when universality is imposed. For arbitrary
$\delta$, we showed the optimal maps are either identity (when
$0\leq \delta \leq .932...$) or the Pauli matrix $\sigma_x$ (when
$.932... \leq \delta \leq \pi/2$).

Lastly, we considered general transformations, i.e.
$(\vartheta,\varphi) \rightarrow (\vartheta-\alpha,\varphi)$, that
are applied only to a partial area of a Bloch sphere ($0\leq
\vartheta \leq \delta$). In a measurement scheme, we considered
four different preparations, $\rho_{1,2}$ and $\sigma_{1,2}$ when
$\alpha=\pi/3$ and $2\pi/3$.  For a quantum scheme, we found that,
in general, the average general transformation yields higher
fidelities than the universal general linear maps studied in
\cite{song1}.  We also found optimal nonlinear general maps for
different values of $\alpha$.

\parindent=0pt
\vspace{6mm}
{\bf Acknowledgments}
\vspace{6mm}

\parindent=6mm
We are grateful to Mark Hillery for helpful discussions. L.H.
acknowledges support from the Royal Society.


\begin{thebibliography}{9}
\bibitem{qc1} A. Ekert and R. Jozsa, Rev. Mod. Phys. {\bf 68}, 733
(1996).
\bibitem{qc2} A.M. Steane, Rep. Prog. Phys. {\bf 61}, 117 (1998).
\bibitem{qc3} V. Vedral and M.B. Plenio, Prog. Quant. Eletron.
{\bf 22}, 1 (1998).
\bibitem{clon1} V. Bu\v{z}ek and M. Hillery, Phys. Rev. A {\bf 54}, 1844 (1996).
\bibitem{clon2} V. Bu\v{z}ek and M. Hillery, Phys. Rev. Lett. {\bf 81}, 5003 (1998).
\bibitem{gisin0} N. Gisin, and S. Massar, Phys. Rev. Lett. {\bf
79}, 2153 (1997).
\bibitem{werner} R.F. Werner, Phys. Rev. A {\bf 58}, 1827 (1998).
\bibitem{bruss} D. Bru\ss, A. Ekert, and C. Michiavello, Phys.
Rev. Lett. {\bf 81}, 2598 (1998).
\bibitem{fuchs} D. Bru\ss, D. DiVincenzo, A. Ekert, C.A. Fuchs, C.
Michiavello, and J.A. Smolin, Phys. Rev. A {\bf 57}, 2368 (1998).
\bibitem{entangler} V. Bu\v{z}ek and M. Hillery, Phys. Rev. A {\bf 62}, 022303 (2000).
\bibitem{disent1} D.R. Terno, Phys. Rev. A {\bf 59}, 3320 (1999).
\bibitem{disent2} T. Mor and D.R. Terno, Phys. Rev. A {\bf 60}, 4341 (1999).
\bibitem{disent3} T. Mor, Phys. Rev. Lett. {\bf 83}, 1451 (1999).
\bibitem{ghosh}  S. Ghosh, S. Bandyopadhyay, A. Roy, D. Sarkar, and G.
Kar, Phys. Rev. A {\bf 61}, 052301 (2000).
\bibitem{hillery} V. Bu\v{z}ek and M. Hillery, Phys. Rev. A {\bf 62}, 052303 (2000).
\bibitem{buzek1} V. Bu\v{z}ek, M. Hillery, and F. Werner, Phys. Rev. A {\bf 60}, R2626 (1999).
\bibitem{buzek2} V. Bu\v{z}ek, M. Hillery, and F. Werner,  J. Mod. Opt. {\bf 47}, 2112 (2000).
\bibitem{gisin} N. Gisin and S. Popescu, Phys. Rev. Lett. {\bf 83}, 432 (1999).
\bibitem{song1} L. Hardy and D.D. Song, Phys. Rev. A {\bf 63}, 032304 (2001).
\bibitem{gisin1} N. Gisin, (in a private conversation).
\bibitem{mea1} S. Massar and S. Popescu, Phys. Rev. Lett. {\bf {74}}, 1259 (1995).
\bibitem{mea2} R. Derka, V. Bu\v{z}ek, and A. Ekert, Phys. Rev.
Lett. {\bf 80}, 1571 (1998).


\end{thebibliography}
\end{document}